\documentclass[prb,letterpaper,aps,floatfix,superscriptaddress,twocolumn]{revtex4-1}
\usepackage{graphicx}
\usepackage{amsmath}
\usepackage{subfigure}
\newcommand{\beq}{\begin{equation}}
\newcommand{\eeq}{\end{equation}}
\newcommand{\bk}{{{\bf{k}}}}

\newcommand{\br}{{{\bf{r}}}}

\newcommand{\bA}{{\bf{A}}}

\newcommand{\bq}{{\bf{q}}}

\newcommand{\bsigma}{{\boldsymbol \sigma}}
\newcommand{\btau}{{\boldsymbol \tau}}
\newcommand{\bnabla}{{\boldsymbol \nabla}}
\newcommand{\beqa}{\begin{eqnarray}}
\newcommand{\eeqa}{\end{eqnarray}}
\newcommand{\pdg}{{\vphantom \dag}}
\newcommand{\dg}{{\dag}} 
\begin{document}
\title{Axion response in Weyl semimetals}
\author{Y. Chen}
\affiliation{Department of Physics and Astronomy, University of Waterloo, Waterloo, Ontario 
N2L 3G1, Canada}
\author{Si Wu}
\affiliation{Department of Physics, University of Toronto, Toronto, Ontario, M5S 1A7, Canada}
\author{A.A. Burkov}
\affiliation{Department of Physics and Astronomy, University of Waterloo, Waterloo, Ontario 
N2L 3G1, Canada}
\date{\today}
\begin{abstract}
Weyl semimetal is a new phase of matter that provides the first solid state realization of chiral 
Weyl fermions. Most of its unique physics is a consequence of chiral anomaly, namely nonconservation 
of the number of particles of a given chirality. Mathematically, this is expressed in the appearance of 
the so called $\theta$-term in the action of the electromagnetic field, when the Weyl fermions are integrated out. 
Recently, however, it has been suggested that the analogy between the chiral fermions of quantum field theory 
 with unbounded linear dispersion, and their solid state realization with a dispersion naturally bounded by the bandwidth
 and crystal momentum defined only within the first Brillouin zone, holds only in a restricted sense, with parts of the 
 $\theta$-term absent. Here we demonstrate that this is not the case. We explicitly derive the $\theta$-term 
 for a microscopic model of a Weyl semimetal by integrating out fermions coupled to electromagnetic field, and show that 
 the result has exactly the same form as in the case of relativistic chiral fermions.   
\end{abstract}
\maketitle
\section{Introduction}
\label{sec:1}
Weyl semimetal is a new proposed phase of matter, which may be thought of as a three-dimensional (3D) analog of 
graphene.~\cite{Volovik,Wan11,Ran11,Burkov11,Xu11}
It is more than ``3D graphene", however, since it possesses interesting topological properties, which are directly tied 
to its three-dimensionality and are not present in graphene. 
These properties may be described as being distinct consequences of a single underlying phenomenon, the 
chiral anomaly.~\cite{Aji12,Son12,Burkov12,Burkov12-2,Spivak12,Grushin12,Goswami12}
Chiral anomaly is a well-known phenomenon in relativistic quantum field theory, which manifests in nonconservation of the 
number of particles of a given chirality in the presence of topologically nontrivial configurations of the background 
gauge field, even when these numbers are conserved classically.~\cite{Adler69,Jackiw69,Nielsen83} 
It is an intrinsically quantum phenomenon with no classical analog and, along with other kinds of quantum anomalies, 
plays an important role in the modern understanding of topologically nontrivial phases of matter.~\cite{Ryu12,Furusaki12,Wen13}

Mathematically, chiral anomaly and related phenomena may be compactly expressed as an induced $\theta$-term in the action of the electromagnetic 
field, when the chiral fermions are integrated out:
\beq
\label{eq:1}
S = \frac{e^2}{32 \pi^2} \int d t d^3 r \,\, \theta({\bf r}, t) \epsilon^{\mu \nu \alpha \beta} F_{\mu \nu} F_{\alpha \beta},
\eeq
where $\theta({\bf r}, t)$ is an ``axion" field,~\cite{Wilczek87} $F_{\mu \nu} = \partial_{\mu} A_{\nu} - \partial_{\nu} A_{\mu}$, 
and we will use $\hbar = c = 1$ units henceforth. 
The spatial and temporal dependence of the axion field defines a specific realization of a topologically-nontrivial 
phase of matter, which can be related to chiral anomaly. 
For example, $\theta = \pi$ in the case of 3D topological insulators.~\cite{Kane10,Qi11}
We have suggested that electromagnetic response of a Weyl semimetal is also 
described by an action of the form of Eq.~\eqref{eq:1}, where the axion field is given by:~\cite{Burkov12,Goswami12}
\beq
\label{eq:2}
\theta({\bf r}, t) = 2 {\bf b} \cdot {\bf r} - 2 b_0 t, 
\eeq
where $2 {\bf b}$ is the separation between the Weyl nodes in momentum space, while $2 b_0$ is the separation between the nodes 
in energy (we will specialize to the case of a Weyl semimetal with only two nodes for simplicity; generalization to any even number of nodes 
is straightforward). 
In Ref.~\onlinecite{Burkov12} we have explicitly derived this term for a generic low-energy model of a Weyl semimetal with an unbounded linear 
dispersion using Fujikawa's method.~\cite{Fujikawa}
This type of $\theta$-term has been discussed before in the context of Lorentz-violating extensions of quantum 
electrodynamics.~\cite{Carroll90,Colladay98,Jackiw99,Volovik05}

It is useful to integrate by parts and rewrite Eq.~\eqref{eq:1} in the following form:
\beq
\label{eq:2.1}
S = - \frac{e^2}{8 \pi^2} \int d^3 r d t \,\,\partial_{\mu} \theta \epsilon^{\mu \nu \alpha \beta} A_{\nu} \partial_{\alpha} A_{\beta}, 
\eeq
which has the form of a ``3D Chern-Simons" term. 
Varying Eq.~\eqref{eq:2.1} with respect to the gauge field, we obtain topological currents, which are the experimentally observable topological 
responses of a Weyl semimetal:
\beqa
\label{eq:3}
j_{\nu}&=&\frac{e^2}{2 \pi^2} b_{\mu} \epsilon^{\mu \nu \alpha \beta} \partial_{\alpha} A_{\beta}, \,\, \mu  = 1, 2, 3, \nonumber \\
j_{\nu}&=&- \frac{e^2}{2 \pi^2} b_0 \epsilon^{0 \nu \alpha \beta} \partial_{\alpha} A_{\beta}. 
\eeqa
The first of Eqs.~\eqref{eq:3} describes the anomalous Hall effect (AHE), associated with the chiral Fermi arc surface states of the Weyl semimetal, 
while the second describes the chiral magnetic effect (CME), which is a current, driven by an applied magnetic field. 

Recently, however, this picture has been challenged in Ref.~\onlinecite{Franz13}, which claims that while the AHE does indeed occur in Weyl 
semimetals, the CME, and the corresponding part of the $\theta$-term in Eq.~\eqref{eq:1}, is an artifact of the unbounded linear dispersion (Dirac sea) 
of relativistic quantum field theory and is thus absent in a solid-state realization of Weyl fermions, with a natural finite cutoff for both momentum and energy. 
This conclusion stands at odds with the idea of topological origin of the chiral anomaly and related phenomena and thus requires a very serious consideration. 
Here we demonstrate that this conclusion is in fact incorrect. We explicitly derive the $\theta$-term by integrating out fermions coupled to electromagnetic 
field in a microscopic model of a Weyl semimetal, introduced by one of us in Ref.~\onlinecite{Burkov11}. We demonstrate explicitly that the $\theta$-term has exactly 
the form of Eq.~\eqref{eq:1}, if allowance is made for the inherent spatial anisotropy of the microscopic model. 
We also show that the origin of the discrepancy between our results and the results of Ref.~\onlinecite{Franz13} is the difference in the order of limits of zero 
frequency and zero wavevector 
when calculating the response function, or transport coefficient,  describing the effect. 
Thus, while the calculation in Ref.~\onlinecite{Franz13} is formally correct, the conclusions, drawn from the result, and its interpretation, are not. 

The rest of the paper is organized as follows. In Section~\ref{sec:2} we provide a detailed derivation of the part of the $\theta$-term, responsible 
for CME. We also discuss the reasons for the apparent discrepancy between our results and the results of Ref.~\onlinecite{Franz13}. 
In Section~\ref{sec:3} we derive the part of the $\theta$-term, responsible for AHE. 
We finish with a brief summary of our results and conclusions in Section~\ref{sec:4}. 
\section{Derivation of the $\theta$-term: CME part}
\label{sec:2}
We start from the microscopic model of a Weyl semimetal in a multilayer heterostructure, made of alternating layers of TI and normal insulator (NI) material, 
introduced in Ref.~\onlinecite{Burkov11}. 
The Hamiltonian is given by:
\beqa
\label{eq:4}
H&=&\int d^3 r \Psi^\dg(\br) \left[v_F \tau^z (\hat z \times \bsigma) \cdot \left(- i \bnabla + e \bA \right) + \hat \Delta(A_z) \right. \nonumber \\
&+&\left.  b \sigma^z + \lambda \tau^y \sigma^z \right] \Psi^\pdg(\br), 
\eeqa
where $v_F$ is the Fermi velocity, characterizing the surface states in TI layers, $\bsigma$ are electron spin operators, 
$\btau$ are pseudospin operators, describing the {\em which-surface} degree of freedom in each TI layer, 
and $\bA$ is the vector potential of the external electromagnetic field. 
$\hat \Delta$ is an operator, describing the tunneling of electrons among neighboring layers in the growth ($z$) direction of the 
heterostructure. Explicitly it is given by:
\beqa
\label{eq:5}
\hat \Delta (A_z)&=&\Delta_S \tau^x \delta_{i,j} + \frac{\Delta_D}{2} \left(\tau^+ e^{i e A_{i z} d} \delta_{j,i+1}\right. \nonumber \\
&+&\left.\tau^- e^{- i e A_{i z} d} \delta_{j, i-1}\right), 
\eeqa
where $\Delta_{S,D}$ are matrix elements, characterizing electron  tunneling between the top and bottom surfaces of the same and nearest-neighbor 
TI layers. We have also assumed that the electromagnetic field is weak enough, so that modification of the tunneling matrix elements by the field 
can be ignored and only the Aharonov-Bohm phase, accumulated over many layers, is nonnegligible. 
The term, proportional to $b$, describes the spin-splitting due to magnetic impurities, which are assumed to be ordered ferromagnetically.
Finally, the last term in Eq.~\eqref{eq:4} describes spin-orbit (SO)-interaction term, induced by broken inversion symmetry in the growth direction of the multilayer. 
This was introduced by us in Ref.~\onlinecite{Burkov12-2} and is needed to create an energy difference between the Weyl nodes, which leads to CME in the 
presence of an external magnetic field. 
\begin{figure}[t]
\centering
\includegraphics[width=8cm]{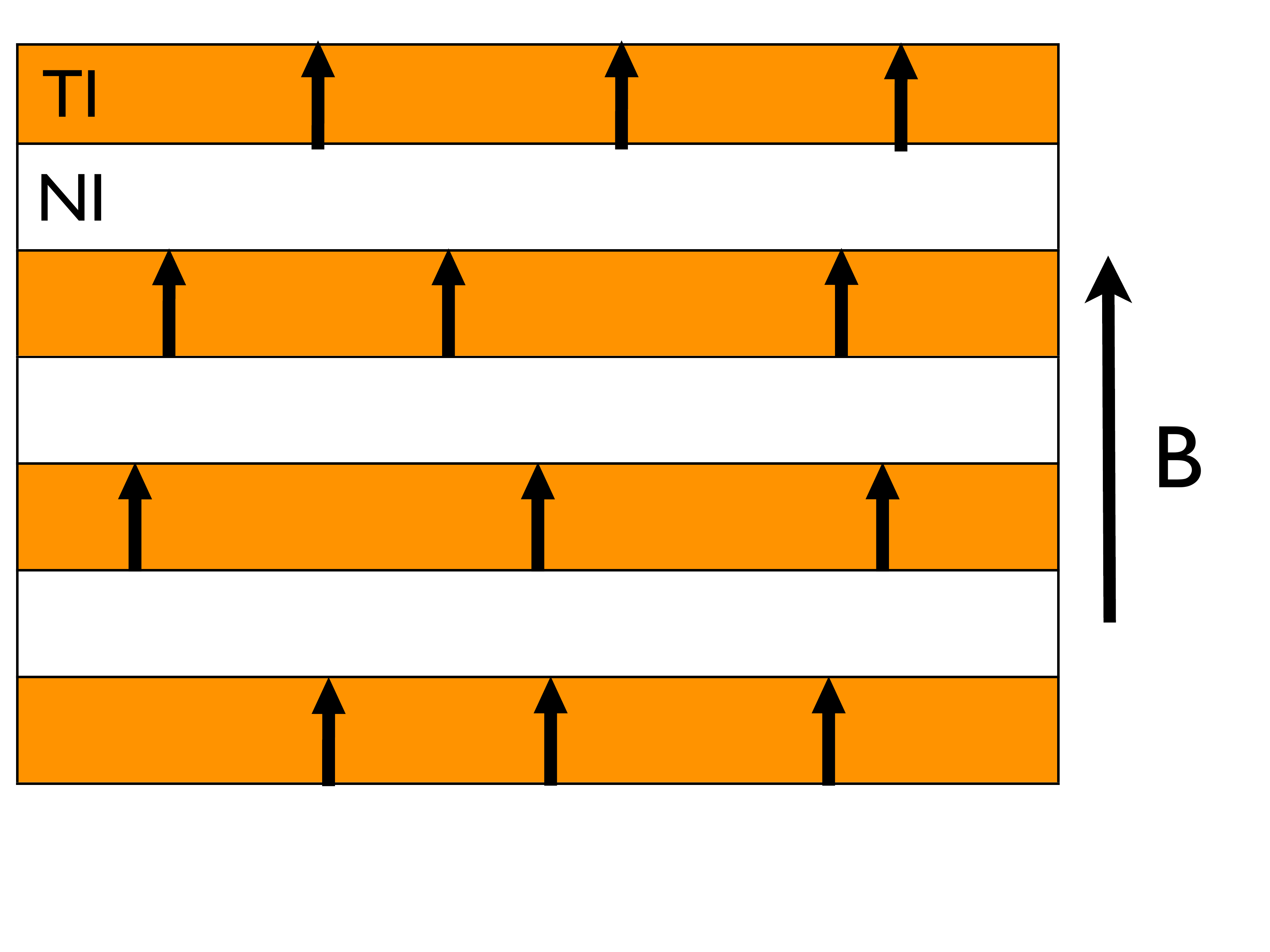}
\caption{(Color online). Schematic picture of the specific realization of Weyl semimetal we consider: a multilayer heterostructure, consisting of 
alternating layers of topological insulator (TI) and normal insulator (NI), doped with magnetic impurities, which are assumed to order ferromagnetically. 
An external magnetic field $B$ is applied along the growth direction of the heterostructure.}
\label{fig:1}
\end{figure}

We now want to integrate out fermions and arrive at an induced action for the electromagnetic field. 
For technical reasons it is convenient to split the calculation into two parts. 
In this section we will focus on evaluating the part of the $\theta$-term, which is responsible for CME, while 
in the following section we will evaluate the part, responsible for AHE. 

We will assume, without loss of generality, that the electromagnetic field consists of a magnetic field in the $z$-direction, 
and a vector potential $A_z$, whose time derivative gives the $z$-component of the electric field $E_z = - \partial_t A_z$. 
We will allow for a time and $z$-coordinate dependence of the vector potential $A_z$ and of the field $\lambda$, but 
assume that the magnetic field is time-independent and uniform. This assumption is made only for technical 
reasons, since it greatly simplifies all calculations. We are ultimately interested in the case of $\lambda$ being time-independent 
and uniform, but a time-dependent magnetic field. We will argue later that, at least to leading order in $\lambda$ and at low frequencies,
the response of the system is identical in both cases. 
We could alternatively do the calculation below treating the time dependence of the magnetic field quasiclassically and assuming a time-independent 
$\lambda$ from the start, but find the way we take below to be a little more transparent. 

It will be convenient to use the Landau level basis of the inversion-symmetric multilayer for this part of the calculation, i.e. the basis of the eigenstates 
of the following Hamiltonian:
\beq
\label{eq:6}
{\cal H}(k_z) = v_F \tau^z (\hat z \times \bsigma) \cdot \left(- i \bnabla + e \bA \right) + \hat \Delta(k_z) + b \sigma^z, 
\eeq
where $\hat \Delta(k_z) = \Delta_S \tau^x + \frac{1}{2}(\Delta_D \tau^+ e^{ i k_z d} + H.c.)$ is the interlayer tunneling 
operator, partially diagonalized by Fourier transform with respect to the layer index and $k_z$ is the corresponding
component of the crystal momentum, defined in the first Brillouin zone (BZ) $(-\pi/d, \pi/d)$ of the multilayer superlattice.
After the canonical transformation $\sigma^{\pm} \rightarrow \tau^z \sigma^{\pm},\,\, \tau^{\pm} \rightarrow \sigma^z \tau^{\pm}$, 
the Hamiltonian takes the form in which the spin and pseudospin degrees of freedom decouple:
\beq
\label{eq:7}
{\cal H}(k_z) =  v_F (\hat z \times \bsigma) \cdot \left(- i \bnabla + e \bA \right) + [b + \hat \Delta(k_z)] \sigma^z. 
\eeq
The tunneling operator can now be diagonalized separately from the rest of the Hamiltonian, which gives:
\beq
\label{eq:8}
{\cal H}_{t}(k_z) =  v_F (\hat z \times \bsigma) \cdot \left(- i \bnabla + e \bA \right) + m_t(k_z)\sigma^z. 
\eeq
Here $t = \pm$ labels the two distinct eigenvalues of the tunneling operator $t \Delta(k_z)$, where 
$\Delta(k_z) = \sqrt{\Delta_S^2 + \Delta_D^2 + 2 \Delta_S \Delta_D \cos(k_z d)}$, and $m_t(k_z) = b + t \Delta(k_z)$. 
The corresponding eigenvectors of the tunneling operator are given by:
\beq
\label{eq:9}
|u^t_{k_z} \rangle = \frac{1}{\sqrt{2}}\left(1, t \frac{\Delta_S + \Delta_D e^{-i k_z d}}{\Delta(k_z)} \right). 
\eeq
The two-component spinor $|u^t_{k_z} \rangle$ is a vector in the $\tau$-pseudospin space.  
To diagonalize the remaining Hamiltonian we pick the Landau gauge $\bA = x B \hat y$.
It is then easily shown~\cite{Burkov12-2} that the eigenstates of ${\cal H}_t(k_z)$ have 
the following form:
\beq
\label{eq:10}
| n, k_y, k_z, s, t \rangle = v^{s t}_{n k_z \uparrow} |n-1, k_y, \uparrow \rangle + v^{s t}_{n k_z \downarrow} | n, k_y, \downarrow \rangle.
\eeq
Here 
\beq
\label{eq:11}
\langle \br | n, k_y, \sigma \rangle = \phi_{n k_y}(\br) |\sigma \rangle, 
\eeq
and $\phi_{n, k_y}(\br)$ are the Landau gauge orbital wavefunctions. 
$s = \pm$ labels the electron-like and hole-like sets of Landau levels:
\beq
\label{eq:12}
\epsilon_{n s t} (k_z) = s \sqrt{2 \omega_B^2 n + m_t^2(k_z)} = s \epsilon_{n t}(k_z), n \geq 1,
\eeq
and the corresponding eigenvectors $|v^{s t}_n\rangle$ are given by:
\beq
\label{eq:13}
|v^{s t}_{n k_z} \rangle = \frac{1}{\sqrt{2}} \left(\sqrt{1 + \frac{m_t(k_z)}{\epsilon_{n t}(k_z)}}, - i s \sqrt{1 - \frac{m_t(k_z)}{\epsilon_{n t}(k_z)}} \right),
\eeq
where $|v^{s t}_{n k_z} \rangle$ is a vector in the $\sigma$-pseudospin space. 
The $n = 0$ Landau level is anomalous, as it is the only Landau level that does not consist of two symmetric electron and hole-like partners. 
Its energy eigenvalues are given by:
\beq
\label{eq:14}
\epsilon_{0 t}(k_z) = - m_t(k_z), 
\eeq
and 
\beq
\label{eq:15}
|v^t_0 \rangle = (0,1). 
\eeq
To simplify the notation we will introduce a composite index $a = (s, t)$ and a tensor product eigenvector:
\beq
\label{eq:16}
|z^a_{n k_z} \rangle = |v^a_{n k_z} \rangle \otimes |u^a_{k_z} \rangle. 
\eeq

We now expand electron field operators in Eq.~\eqref{eq:4} in terms of the complete set of states we have just constructed:
\beq
\label{eq:17}
\Psi^\dg(\br \sigma \tau) = \sum_{n,k_y,k_z} \langle n, k_y, k_z, a | \br, \sigma, \tau \rangle c^\dg_{n k_y k_z a},
\eeq
where summation over repeated $a$ indices will be implicit henceforth, and obtain imaginary time action for our system in Matsubara frequency space in the following form:
\beqa
\label{eq:18}
&&S = \sum_{n, k_y, k_z, i \omega} {\cal G}^{-1}_{n k_y a}(k_z, i \omega) c^\dg_{n k_y k_z a}(i \omega) c^\pdg_{n k_y k_z a}(i \omega) \nonumber \\
&+&\sum_{n, k_y, k_z, k_z'} \sum_{i \omega, i \omega'} \delta {\cal G}^{-1}_{n k_y a a'}(k_z,k_z' | i\omega - i \omega') \nonumber \\
&\times&c^\dg_{n k_y  k_z a}(i \omega) c^\pdg_{n k_y k_z' a'}(i \omega'),\nonumber \\
\eeqa
where 
\beqa
\label{eq:19}
&&{\cal G}^{-1}_{n k_y  a}(k_z, i \omega) = - i \omega - \mu + \epsilon_{n a}(k_z), \nonumber \\
&&\delta {\cal G}^{-1}_{n k_y a a'}(k_z, k_z' | i\omega - i \omega') = \nonumber \\
&&\frac{1}{\sqrt{V \beta}} \lambda(k_z - k_z', i \omega - i \omega') \langle z^a_{n k_z}| \tau^y | z^{a'}_{n k_z'} \rangle \nonumber \\
&-&\frac{1}{\sqrt{V \beta}} A_z(k_z - k_z', i \omega - i \omega') \langle z^a_{n k_z}| \hat j_z(k_z') | z^{a'}_{n k_z'} \rangle.
\eeqa 
The current operator $\hat j_z$ is obtained by expanding Eq.~\eqref{eq:5} to first order in $A_z$:
\beq
\label{eq:20}
\hat j_z(k_z) = e \Delta_D d \sigma^z \left[\tau^x \sin(k_z d) + \tau^y \cos(k_z d) \right]. 
\eeq
Integrating out fermions perturbatively in $\lambda$ and $A_z$ we obtain at second order the following imaginary time action for the electromagnetic field:
\beq
\label{eq:21}
S = \frac{1}{2} \textrm{Tr} \sum_{i \omega, i \omega'} {\cal G}(i \omega) \delta {\cal G}^{-1}(i \omega - i \omega') {\cal G}(i \omega') \delta {\cal G}^{-1}(i \omega' - i \omega). 
\eeq
The topological term, of interest to us, is proportional to the product of $\lambda$ and $A_z$. 
Leaving only this term in the imaginary time action, performing summation over the Landau level orbital index $k_y$ and 
over the fermion Matsubara frequency, we obtain:
\beq
\label{eq:21.1}
S = B \sum_{q, i \Omega} \Pi(q, i\Omega) A_z(q, i\Omega) \lambda(-q, -i \Omega)
\eeq
where the response function $\Pi(q, i\Omega)$ is given by:
\beqa
\label{eq:22}
\Pi(q, i\Omega)&=&\frac{e}{2 \pi L_z} \sum_{n, k_z} \frac{n_F[\xi_{n a'}(k_z)] - n_F[\xi_{n a}(k_z + q)]}{i \Omega + \xi_{n a'}(k_z) - \xi_{n a}(k_z + q)} \nonumber \\
&\times& \langle z^a_{n k_z}| \hat j_z(k_z) | z^{a'}_{n k_z} \rangle \langle z^{a'}_{n k_z} | \tau^y |z^a_{n k_z} \rangle. 
\eeqa
Here $n_F$ is the Fermi-Dirac distribution function, $\xi_{n a}(k_z) = \epsilon_{n a}(k_z) - \mu$, and the magnetic field $B$ in Eq.~\eqref{eq:21.1} arises 
from the Landau level orbital degeneracy as:
\beq
\label{eq:22.1}
\frac{L_x L_y}{2 \pi \ell_B^2} = \frac{L_x L_y e B}{2 \pi},
\eeq
$\ell_B = 1/ \sqrt{e B}$ being the magnetic length. 
We have also ignored the $q$-dependence of the matrix elements in Eq.~\eqref{eq:22}, which is not important for small $q$. 

At this point we will specialize to the case of an undoped Weyl semimetal, i.e. set $\mu = 0$. 
Then it is clear from Eq.~\eqref{eq:22} that for Landau levels with $n \geq 1$, only terms with $s \neq s'$ contribute due to the difference of Fermi factors 
in the numerator. 
We are interested ultimately in the zero frequency and zero wavevector limit of the response function $\Pi(q, i\Omega)$. 
As is often the case, the value of $\Pi(0,0)$ depends on the order in which the zero frequency and zero wavevector 
limits are taken.
Below we will consider both possibilities separately and discuss their physical meaning. 
We will argue that this difference in the order of limits is precisely the origin of the discrepancy between our 
results for CME and the results of Ref.~\onlinecite{Franz13}. 

Before we proceed with explicit evaluation of the $q \rightarrow 0$ and $\Omega \rightarrow 0$ limit, let us note 
a crucial property of the response function $\Pi(q, \Omega)$.  
If we take into account the following symmetry properties of the matrix elements in Eq.~\eqref{eq:22}:
\beqa
\label{eq:23}
&&\langle v^{+ t}_{n k_z} | v^{- t'}_{n k_z} \rangle  = - \langle v^{+ t'}_{n k_z} | v^{- t}_{n k_z} \rangle, \nonumber \\
&&\langle v^{+ t}_{n k_z} | \sigma^z| v^{- t'}_{n k_z} \rangle  =  \langle v^{+ t'}_{n k_z} | \sigma^z| v^{- t}_{n k_z} \rangle, 
\eeqa
it is easy to see that when the limit $\Omega \rightarrow 0$ is taken, independently of the value of $q$, the $n \geq 1$ Landau levels in fact 
do not contribute at all, mutually cancelling due to Eq.~\eqref{eq:23}. It follows that $\Pi(q, i\Omega)$ at small $\Omega$ is
determined completely by the contribution of the two $n = 0$ Landau levels, whose energy eigenvalues and the corresponding 
eigenvectors are independent of the magnetic field. 
This means that in the small $\Omega$ limit $\Pi(q, i \Omega)$ becomes independent of the magnetic field and the effective action in Eq.~\eqref{eq:21.1} 
then depends linearly on both $\lambda$ and $B$, independently of the magnitude of $B$. 
This in turn implies that in the low-frequency limit the response of the system to fluctuating field $\lambda$ at fixed $B$ is identical to its 
response to fluctuating $B$ at fixed $\lambda$ (this is analogous to Onsager's reciprocity relations for transport coefficients):
\beq
\label{eq:23.1}
 S = \lambda \sum_{q, i \Omega} \Pi(q, i\Omega) A_z(q, i\Omega) B(-q, -i \Omega),
\eeq
which is precisely what we are interested in. 
To double-check the correctness of this argument we have explicitly recalculated the $\theta$-term 
by a different method, namely by treating the magnetic field perturbatively, expanding the 
effective action to third order, and evaluating the coefficient of the term,  proportional 
to the product of $\lambda$, $A_z$ and $\partial_x A_y$, and obtained identical results.  
Same result is also obtained by yet another method, already mentioned above, i.e. by treating 
the time dependence of the magnetic field quasiclassically.

Let us now proceed to explicitly evaluate $\Pi(0,0)$, which determines the low-frequency and long-wavelength 
response of our system. 
Let us first look at the situation when we send $q$ to zero before sending $\Omega$ to zero. 
Explicitly evaluating the matrix elements in Eq.~\eqref{eq:22}, we obtain:
\beqa
\label{eq:24}
\Pi(0,0)&=&- \frac{e \Delta_D d}{2 \pi \ell_B^2} \int_{-\pi/d}^{\pi/d} \frac{d k_z}{2 \pi} n_F[b - \Delta(k_z)] \nonumber \\
&\times&\frac{\Delta_S \Delta_D \sin^2(k_z d) + \Delta^2(k_z) \cos(k_z d)}{\Delta^3(k_z)}. \nonumber \\
\eeqa
The function $n_F[b - \Delta(k_z)]$, assuming zero temperature, restricts the integral over $k_z$ to the interval where $b < \Delta(k_z)$ , i.e. the 
part of the 1D BZ outside of the interval between the Weyl nodes, which assuming both $\Delta_S$ and $\Delta_D$ are positive for concreteness, 
occur at points $\pi/d \pm k_0$, where:~\cite{Burkov12-2}
\beq
\label{eq:25}
k_0 = \frac{1}{d} \arccos\left(\frac{\Delta_S^2 + \Delta_D^2 - b^2}{2 \Delta_S \Delta_D} \right). 
\eeq
Using the identity:
\beqa
\label{eq:26}
&&\int_{-\pi/d}^{\pi/d} \frac{d k_z}{2 \pi} \frac{\Delta_S \Delta_D \sin^2(k_z d) + \Delta^2(k_z) \cos(k_z d)}{\Delta^3(k_z)} \nonumber \\
&=&- \frac{1}{d^2 \Delta_S \Delta_D} \int_{-\pi/d}^{\pi/d} \frac{d k_z}{2 \pi} \frac{d^2 \Delta(k_z)}{d k_z^2} = 0, 
\eeqa
we can, however, rewrite the integral over $k_z$ in Eq.~\eqref{eq:24} in a more intuitive form, as an integral over the interval
between the Weyl nodes:
\beqa
\label{eq:27}
\Pi(0,0)&=&- \frac{e \Delta_D d}{2 \pi \ell_B^2} \int_{\pi/d-k_0}^{\pi/d+k_0} \frac{d k_z}{2 \pi}  \nonumber \\
&\times&\frac{\Delta_S \Delta_D \sin^2(k_z d) + \Delta^2(k_z) \cos(k_z d)}{\Delta^3(k_z)}. \nonumber \\
\eeqa
Evaluating the integral over $k_z$ explicitly, we finally obtain:
\beq
\label{eq:28} 
\lambda \Pi(0,0) = - \frac{e^2 \Delta \epsilon}{4 \pi^2}, 
\eeq
where 
\beq
\label{eq:29}
\Delta \epsilon = \frac{\lambda}{\Delta_S b} \sqrt{[(\Delta_S + \Delta_D)^2 - b^2][b^2 - (\Delta_S - \Delta_D)^2]}, 
\eeq
is the energy difference between the Weyl nodes, induced by the field $\lambda$, as was shown in Ref.~\onlinecite{Burkov12-2}. 
Thus, after Wick's rotation $\tau \rightarrow i t$, $\Delta \epsilon \rightarrow  - i \Delta \epsilon$, we finally obtain the following result for the electromagnetic field action: 
\beq
\label{eq:30}
S =  - \frac{e^2 \Delta \epsilon}{4 \pi^2} \int d^3 r d t \,\, A_z(\br, t) B(\br, t),
\eeq
which has the same form as Eq.~\eqref{eq:2.1}. 
Functional derivative of Eq.~\eqref{eq:30} with respect to $A_z$ gives the current that flows in response to magnetic 
field, i.e. the CME:
\beq
\label{eq:31}
j_z  = - \frac{e^2 \Delta \epsilon}{4 \pi^2} B. 
\eeq
This demonstrates that CME is not an artifact of a low-energy 
model of a Weyl semimetal with an unbounded linear dispersion and exists in a microscopic model just as well. 
This is in agreement with the viewpoint that chiral anomaly has topological origin, independent of details of the dispersion. 

Why was a zero result obtained for CME in Ref.~\onlinecite{Franz13}? 
As already mentioned above, this is an issue of the order of limits when calculating $\Pi(0,0)$. 
Let us now set $\Omega$ to zero before taking the limit $q \rightarrow 0$. 
In this case, in addition to the contribution to $\Pi(0,0)$, given by Eq.~\eqref{eq:27}, which arises due to transitions 
between the $t = +$ and $t = -$ lowest ($n = 0$) Landau levels, there is an extra contribution due to the intra-Landau-level 
processes within the $t = -$ Landau level, which crosses the Fermi energy at the location of the Weyl nodes. 
This extra contribution is given by:
\beqa
\label{eq:31.2}
&&\tilde \Pi(0,0) = \frac{ e}{2 \pi L_z} \sum_{k_z} \left. \frac{d n_F(\epsilon)}{d \epsilon} \right|_{\epsilon = - m_-(k_z)}  \nonumber \\
&\times&\langle z^-_{0 k_z} | \hat j_z(k_z) | z^-_{0 k_z} \rangle \langle z^-_{0 k_z} | \tau^y | z^-_{0 k_z} \rangle, 
\eeqa 
and is easily shown to be equal to Eq.~\eqref{eq:27} in magnitude, but opposite in sign, which means that in this case $\Pi(0,0)$ vanishes. 
Thus, not unexpectedly, the final result for  $\Pi(0,0)$ depends on the order in which the $q \rightarrow 0$ and $\Omega \rightarrow 0$ limits 
are taken:
\beqa
\label{eq:31.3}
&&\lim_{\Omega \rightarrow 0} \lim_{q \rightarrow 0} \lambda  \Pi(q, \Omega) = - \frac{e^2 \Delta \epsilon}{4 \pi^2}, \nonumber \\
&&\lim_{q \rightarrow 0} \lim_{\Omega \rightarrow 0} \lambda \Pi(q, \Omega) = 0. 
\eeqa
What is the physical meaning of these two distinct orders of limits in calculating $\Pi(0,0)$? 
When $q$ is taken to zero first, one is calculating the low-frequency limit of response to 
a time-dependent external field, in our case magnetic field along the $z$-direction. 
This response is finite and represents CME, described by Eq.~\eqref{eq:31}. 
If one takes $\Omega$ to zero first, however, one is calculating a thermodynamic property, in our 
case change of the ground state energy of the system in the presence of an additional static vector potential  
in the $z$-direction:
\beqa
\label{eq:31.4}
j_z&=&\frac{1}{V} \frac{\partial E(A_z)}{\partial A_z} = 0. 
\eeqa
This could be nonzero in, for example, a current-carrying state of a superconductor, which 
possesses phase rigidity, but vanishes identically in our case. 
Thus the issue is the correct interpretation of what CME actually is. 
Our calculation demonstrates that it can not be thought of as persistent ground state current in 
the presence of a static magnetic field (we incorrectly called the CME current ``persistent current" in Refs.~\onlinecite{Burkov12,Burkov12-2}), 
but should rather be thought of as a DC limit of response to a time-dependent magnetic field. 
\section{Derivation of the $\theta$-term: AHE part}
\label{sec:3}
While there is at present no doubt that the AHE, associated with the Weyl nodes, does exist, we will still 
provide a derivation of the corresponding part of the $\theta$-term in Eq.~\eqref{eq:1}, if only for completeness
purposes. 
This is done most conveniently in the crystal momentum basis, rather than the Landau level basis, 
used in the previous section. 
We start from the Hamiltonian Eq.~\eqref{eq:8}, but in the absence of magnetic filed, in which case all components of the 
crystal momentum are conserved:
\beq
\label{eq:32}
{\cal H}_{t}(\bk) =  v_F (\hat z \times \bsigma) \cdot \bk + m_t(k_z)\sigma^z. 
\eeq
Its eigenstate energies and the corresponding eigenvectors are given by:
\beq
\label{eq:33}
\epsilon_{s t}(\bk) = s \sqrt{v_F^2 (k_x^2 + k_y^2) + m_t^2(k_z)} = s \epsilon_{t}(\bk),
\eeq
and 
\beq
\label{eq:34}
|v^{st}_{\bk}\rangle = \frac{1}{\sqrt{2}} \left[ \sqrt{1 + s \frac{m_t(k_z)}{\epsilon_t(\bk)}}, - i s \hat k_+ \sqrt{1 - s \frac{m_t(k_z)}{\epsilon_t(\bk)}} \right]. 
\eeq
Introducing the combined spin and pseudospin eigenvectors, $| z^{st}_\bk \rangle = |v^{st}_\bk \rangle \otimes |u^t_{k_z} \rangle$, we can write 
the imaginary time action of the electrons coupled to electromagnetic field in the form, analogous to Eq.~\eqref{eq:18} in the previous section:
\beqa
\label{eq:35}
S&=&\sum_{\bk,i\omega} {\cal G}^{-1}_{s t}(\bk, i\omega) c^\dg_{s t \bk}(i \omega) c^\pdg_{s t \bk} \nonumber \\
&+&\sum_{\bk, \bk', i\omega, i\omega'} \delta {\cal G}^{-1}_{st, s't'}(\bk, \bk'|i \omega - i \omega') c^\dg_{s t \bk}(i\omega) c^\pdg_{s't' \bk'}(i \omega'), \nonumber \\
\eeqa
where summation over the $s,t$ indices has been made implicit and:
\beqa
\label{eq:36}
&&{\cal G}^{-1}_{s t}(\bk, i\omega) = - i \omega + \xi_{s t \bk}, \nonumber \\
&&\delta {\cal G}^{-1}_{s t, s' t'}(\bk, \bk' | i\omega - i \omega') = \frac{e v_F}{\sqrt{V \beta}} \langle z^{s t}_\bk | \sigma^x | z^{s ' t'}_{\bk'} \rangle \nonumber \\
&\times& A_y(\bk - \bk', i \omega - i \omega') + \frac{i e}{\sqrt{V \beta}} \langle z^{s t}_{\bk} | z^{s' t'}_{\bk'} \rangle \nonumber \\
&\times&A_0(\bk - \bk', i\omega - i \omega'), 
\eeqa
where we have employed Landau gauge for the vector potential $\bA = x B \hat y$ and have also included the temporal component 
of the gauge field $A_0$. 
Integrating out fermions, we again obtain, at second order in the gauge fields, an effective action for the electromagnetic field of the 
form of Eq.~\eqref{eq:21}. 
As before, we are interested only in the topological part of this action, which is proportional to the product of $A_0$ and $A_y$ and is given by:
\beqa
\label{eq:37}
S&=&\frac{i e^2 v_F}{V} \sum_{\bk, \bq, i\Omega} \frac{n_F(\xi_{s' t' \bk}) - n_F(\xi_{s t \bk + \bq})}{i \Omega + \xi_{s' t' \bk} - \xi_{s t \bk + \bq}} 
\langle z^{s t}_{\bk + \bq} | z^{s' t'}_{\bk} \rangle \nonumber \\
&\times&\langle z^{s' t'}_{\bk} | \sigma^x | z^{s t}_{\bk + \bq} \rangle A_0(\bq, i\Omega) A_y(-\bq, -i\Omega). 
\eeqa   
At this point we again specialize to the case of an undoped Weyl semimetal with $\mu =0$. In this case we must have $s \neq s'$, since 
the $s = s'$ contributions in Eq.~\eqref{eq:37} vanish due to the difference of Fermi-Dirac distribution functions in the numerator. 
We then set $\Omega = 0$ in the denominator and assume, for simplicity, that the only spatial dependence of the gauge fields comes from the 
Landau gauge choice, which implies that $\bq = q \hat x$ (we will comment on the order of limits issue below). 
We now note that, among all the distinct factors in Eq.~\eqref{eq:37}, $\langle z^{+t}_{\bk + \bq} | z^{- t}_{\bk} \rangle \rightarrow 0$, when $\bq \rightarrow 0$, 
while all others remain finite in this limit. 
Thus, to leading nontrivial order in $\bq$, we can expand $\langle z^{+t}_{\bk + \bq} | z^{- t}_{\bk} \rangle$ to first order in $\bq$, while setting $\bq \rightarrow 0$ everywhere  
else. 
We obtain:
\beq
\label{eq:38}
\langle z^{+t}_{\bk + \bq} | z^{- t}_{\bk} \rangle \approx - \frac{v _F q}{2 \epsilon_t(\bk) \sqrt{k_x^2 + k_y^2}} \left[i k_y + \frac{m_t(k_z) k_x}{\epsilon_t(\bk)}\right]. 
\eeq
Substituting this into Eq.~\eqref{eq:37}, we obtain, after straightforward algebra:
\beq
\label{eq:39}
S = \frac{e^2 v_F^2}{2 V} \sum_{\bk, \bq, i\Omega,t} \frac{m_t(k_z)}{\epsilon^3_t(\bk)} A_0(\bq, i \Omega) q A_y(-\bq, -i \Omega). 
\eeq
Note that the quantity 
\beq
\label{eq:39.1}
\Omega^z_{t \bk} = \frac{v_F^2 m_t(k_z)}{2 \epsilon^3_t(\bk)}, 
\eeq
which appears in Eq.~\eqref{eq:39}, has the meaning of the $z$-component of the Berry curvature of the two filled bands. 
Performing the integrals over $k_x, k_y$, we obtain:
\beqa
\label{eq:40}
S&=&\frac{i e^2}{4 \pi} \sum_{\bq, i\Omega,t} \int^{\pi/d}_{-\pi/d} \frac{d k_z}{2 \pi} \textrm{sign}[m_t(k_z)] A_0(\bq, i\Omega) \nonumber \\
&\times&(-i  q) A_y(-\bq, -i\Omega). 
\eeqa
The $\textrm{sign}[m_t(k_z)]$ function in Eq.~\eqref{eq:40} expresses the fact that the Berry flux along the $z$-direction changes sign at the 
location of the Weyl nodes, which are monopole sources of the Berry curvature. 
Integration over $k_z$, which simply gives the distance in momentum space between the Weyl nodes, and Wick's rotation to real time $\tau \rightarrow i t$, finally gives:
\beq
\label{eq:41}
S = - \frac{e^2 k_0}{2 \pi^2} \int d^3 r d t \,\, A_0(\br, t) \partial_x A_y(\br, t), 
\eeq
which has exactly the form of Eq.~\eqref{eq:2.1}. 
Functional derivative of $S$ with respect to $A_y$ gives the Hall current in response to the electric field:
\beq
\label{eq:42}
j_y = - \frac{e^2 k_0}{2 \pi^2} \partial_x A_0 = - \frac{e^2 k_0}{2 \pi^2} E_x, 
\eeq
which gives the correct expression for the anomalous Hall conductivity of Weyl semimetal.~\cite{Burkov11}
Note that in this calculation, the issue of the order of limits, which was important in the calculation of CME, never arises. 
The reason for this is that while CME is associated with bulk extended states, AHE is due to the Fermi arc surface states, 
arising from completely filled bulk states, which makes the order of limits irrelevant. 
\section{Conclusions}
\label{sec:4}
The goal of this paper was to demonstrate explicitly that the $\theta$-term topological response of Weyl semimetals, described 
by Eq.~\eqref{eq:1} is indeed ``topological", in the sense that it does not depend on details of the bandstructure. 
Our previous derivation of the $\theta$-term was criticized in Ref.~\onlinecite{Franz13} as being based on a generic low-energy 
model of a Weyl semimetal with an unbounded linear dispersion. 
We have shown here that exactly the same result is obtained in a specific microscopic model of a Weyl semimetal, with 
no cutoff and regularization issues ever arising. 
Our calculations have also clarified the physical meaning of CME, --- one of the phenomena associated with 
the $\theta$-term in Weyl semimetals. 
We have shown that CME should not be thought of as persistent current in response to static magnetic field, but rather 
as a DC limit of response to a time-dependent magnetic field. 
Our work, along with related recent work on this subject,~\cite{Grushin12,Goswami12} demonstrates unambiguously
that Weyl semimetals provide the first realization of ``axion electrodynamics".~\cite{Wilczek87} 
An interesting extension of the above calculation is to include the effect of electron-electron interactions, in particular
investigating the interplay of chiral anomaly (which by itself is presumably insensitive to interactions) with low-energy and long-wavelength collective modes of a Weyl semimetal. 
This will be reported elsewhere. 
\begin{acknowledgments}
We acknowledge useful discussions with A.~Paramekanti. Financial support was provided by the NSERC of Canada and a University of Waterloo start up grant. 
\end{acknowledgments}

\end{document}